\documentclass[aip,apl,amsmath,amssymb,reprint,]{revtex4-1}
\usepackage{graphicx}
\usepackage{epstopdf}
\usepackage{dcolumn}
\usepackage{bm}
\usepackage{color}

\begin{document}

\title{Optically Addressed Spatial Light Modulator based on Nonlinear Metasurface}

\author{Shengchao Gong}
\affiliation{The Key Laboratory of Weak-Light Nonlinear Photonics, Ministry of Education, School of Physics and TEDA Applied Physics Institute, Nankai University, Tianjin, 300071, P.R. China}

\author{Mengxin Ren}
\email{ren$_$mengxin@nankai.edu.cn}
\affiliation{The Key Laboratory of Weak-Light Nonlinear Photonics, Ministry of Education, School of Physics and TEDA Applied Physics Institute, Nankai University, Tianjin, 300071, P.R. China}
\affiliation{Collaborative Innovation Center of Extreme Optics, Shanxi University, Taiyuan, Shanxi 030006, P.R. China}
\affiliation{Renewable Energy Conversion and Storage Center, Nankai University, Tianjin 300071, P.R. China}

\author{Wei Wu}
\affiliation{The Key Laboratory of Weak-Light Nonlinear Photonics, Ministry of Education, School of Physics and TEDA Applied Physics Institute, Nankai University, Tianjin, 300071, P.R. China}

\author{Wei Cai}
\affiliation{The Key Laboratory of Weak-Light Nonlinear Photonics, Ministry of Education, School of Physics and TEDA Applied Physics Institute, Nankai University, Tianjin, 300071, P.R. China}

\author{Jingjun Xu}
\email{jjxu@nankai.edu.cn}
\affiliation{The Key Laboratory of Weak-Light Nonlinear Photonics, Ministry of Education, School of Physics and TEDA Applied Physics Institute, Nankai University, Tianjin, 300071, P.R. China}

\begin{abstract}
Spatial light modulators (SLMs) are devices for modulating amplitude, phase or polarization of a light beam on demand. Such devices have been playing an indispensable influence in many areas from our daily entertainments to scientific researches. In the past decades, the SLMs have been mainly operated in electrical addressing (EASLM) manner, wherein the writing images are created and loaded via conventional electronic interfaces. However, adoption of pixelated electrodes puts limits on both resolution and efficiency of the EASLMs. Here, we present an optically addressed SLM based on a nonlinear metasurface (MS-OASLM), by which signal light is directly modulated by another writing beam requiring no electrode. The MS-OASLM shows unprecedented compactness and is 400~nm in total thickness benefitting from the outstanding nonlinearity of the metasurface. And their subwavelength feature size enables a high resolution up to 250 line pairs per millimeter, which is more than one order of magnitude better than any currently commercial SLMs. Such MS-OASLMs could provide opportunities to develop the next generation of high resolution displays and all-optical information processing technologies.
\end{abstract}
\maketitle

More than forty years ago, enthusiasm for large-screen display and optical data processing in parallel architectures promoted invention and development of spatial light modulators (SLMs).\cite{goodman1968Introduction,heynick1973special,casasent1978optical,tanguay1983spatial} Such devices can manipulate light by modulating its amplitude, phase or polarization in dimensions of space and time as demanded.\cite{Efron1994Spatial,savage2009digital} Nowadays, the SLMs already make its quiet, yet steady entrance into our daily lives and cutting edge researches, ranging from portable displays, virtual reality, light detection and ranging, and adaptive optics, etc.\cite{vettese2010microdisplays,maimone2017holographic,christian2013survey,harke2008resolution,barredo2016atom,chiou2005massively,beckers1993adaptive,katz2012looking} Depending on the way that information is written into the SLMs, the SLMs can be divided into electrically addressed (EASLMs) and optically addressed (OASLMs).\cite{moddel1997spatial}

The EA is the most mature and widely used technique in currently commercial SLMs. The EASLMs are commonly constructed by sandwiching a light modulating material, typically liquid crystals (LCs, as shown in Fig.~1) between pixelated electrical addressing electrodes. A voltage is applied across to tilt the LC molecules, providing means for a dynamical control of light such as phase, polarization state, and intensity, as shown in Fig.~1.\cite{yang2014fundamentals} However, pitches of the electrode pixels are normally on micrometers level even fabricated by state-of-the-art CMOS foundry. Thus the electrode meshes act as two-dimensional gratings, and induce undesirable diffraction and wavefront distortions to the readout beam, which not only suppress light utilization efficiency, but also introduce noise to the optical displaying and the data processing. Furthermore, dead spaces between the pixels significantly block the light and reduce brightness of the SLMs.

In contrast, in the OASLMs, light modulation is fulfilled in an all-optical manner by irradiating a writing light beam rather than applying the electronic signal.\cite{margerum1970reversible,white1970liquid} Thus no physically discrete electrode pixel is required, which shows advantages of easy fabrication, no diffraction, and a nearly 100\% fill factor (fraction of an area that is optically usable).\cite{shrestha2015high,miri2014Liquid} Such OASLMs are highly desirable, because they enable light control directly by light without electronic-optical conversion as EASLMs do. Thus they allow unique all-optical applications that are impossible by EASLMs, including coherent to incoherent image conversion, wavelength conversion, real time optical correlation, and parallel optical computation.\cite{smith1985lasers,zhang1997incoherent,shih2001all,woods2012optical,solodar2016infrared}

\begin{figure*}[htpb]
\includegraphics[width=180mm]{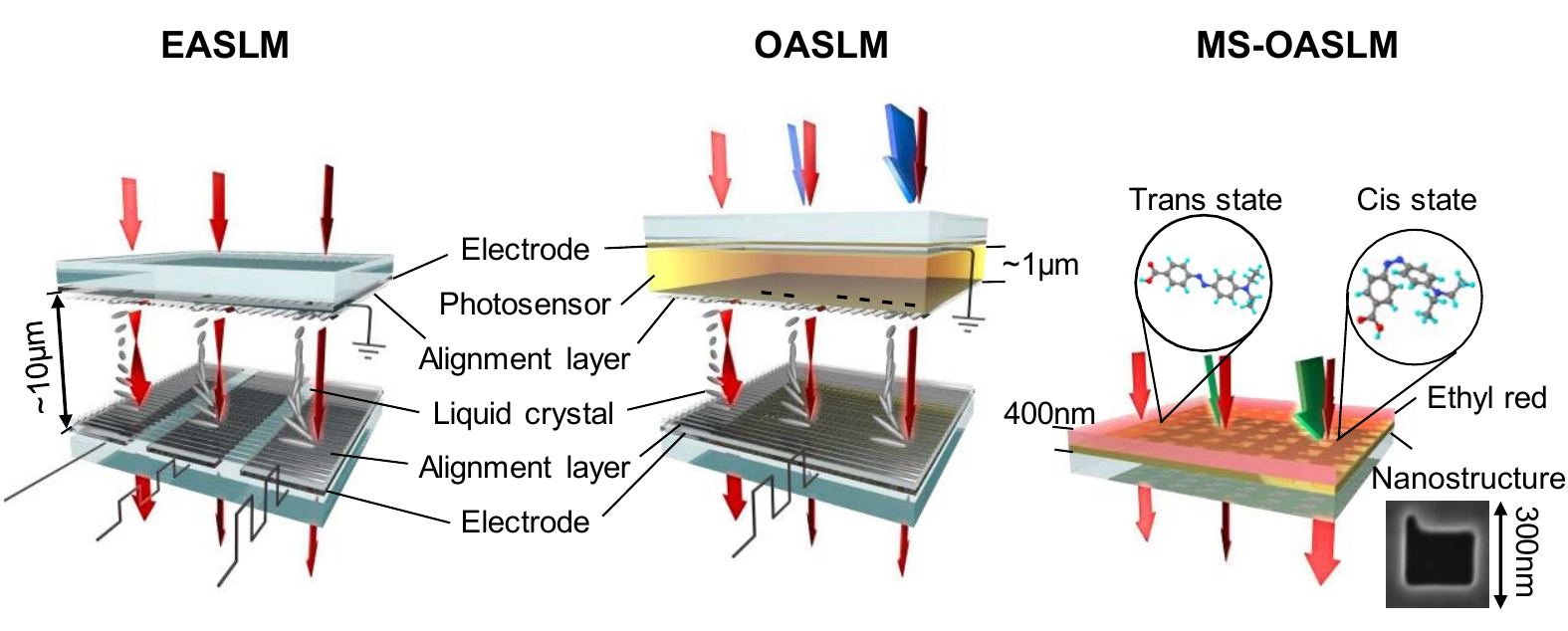}\caption{\label{fig1}
\textbf{Comparison between different SLMs.} EASLM (left): pixels are addressed electrically. A voltage is applied across the LCs through pixelated electrodes, leading to a certain tilt of the LC molecules. Resulting from optical anisotropy in the LC molecules, this tilt would modify phase, polarization (in picture), or intensity (aided by extra polarizers) of light (red arrows). OASLM (middle): light properties are controlled directly by another write beam (usually using violet light, shown by blue arrows). The device consists of a photosensor followed by a LC layer, which are sandwiched between two uniform transparent electrodes. The photosensor is locally discharged by the incident light and its resistance reduces. Thereby the LCs rotate under an external voltage between the electrodes, leading to a modulation to the signal light. MS-OASLM (right): metasurface is a sole constructing component, in which both electrodes and photosensor are absent. The writing beam (green arrows) causes a spatially heterogeneous photoisomerization of ethyl red azo molecules and selectively tunes the plasmonic resonance of the metasurface. This in turn affects the readout light polarization in a nonlinear manner. Such MS-OASLM shows superiorities of ultracompactness and easy fabrication. Inset gives a SEM image of the metasurface unit cell.}
\end{figure*}

In principle, the OASLMs can be made using a single nonlinear optical material. For example, single photorefractive materials were used to realize updatable holographic display, image amplifier, phase conjugator.\cite{xu1995holographic,grunnet1997spontaneous,hesselink1998photorefractive,goonesekera2000image,yaqoob2008optical,tay2008updatable} But the devices were too bulky and inapplicable in the nano-era because of the too weak nonlinearity. During the past decades, researchers never gave up but never succeeded in finding a proper material with giant nonlinearity, which supports efficient ``light-control-by-light'' operation within a small-volume. Thus an OASLM monolithically based on a nonlinear material still remains an illusion. As a compromised strategy, a hybrid OASLM configuration consisting of light modulating layer (again mostly LCs), photosensing layer (such as ZnO, CdS, a-Si, or AsS, \cite {white1970liquid,beard1973ac,moddel1989high,fukushima1990bistable,miri2014Liquid}) and electrodes were proposed decades ago, as shown in Fig.~1.\cite {margerum1970reversible,white1970liquid,moddel1989high,shrestha2015high,solodar2019highly} The photosensitive layer absorbs the incident light and locally discharges, leading to a reduced resistance. LCs tilt under the voltage between the electrodes, and finally the signal light is modulated. In order to weaken the writing light, a thicker photosensor is preferred, but it limits the resolution due to lateral charge diffusion in the photosensing layer, and also compromises the device compactness.\cite{li1995resolution} Moreover, the operations of such OASLMs critically rely on precise opto-electronic synchronism between the write/read optical pulses and the external voltage bursts, which require complicated clocking circuits.\cite{moddel1989high}

\begin{figure*}[htpb]
\includegraphics[width=110mm]{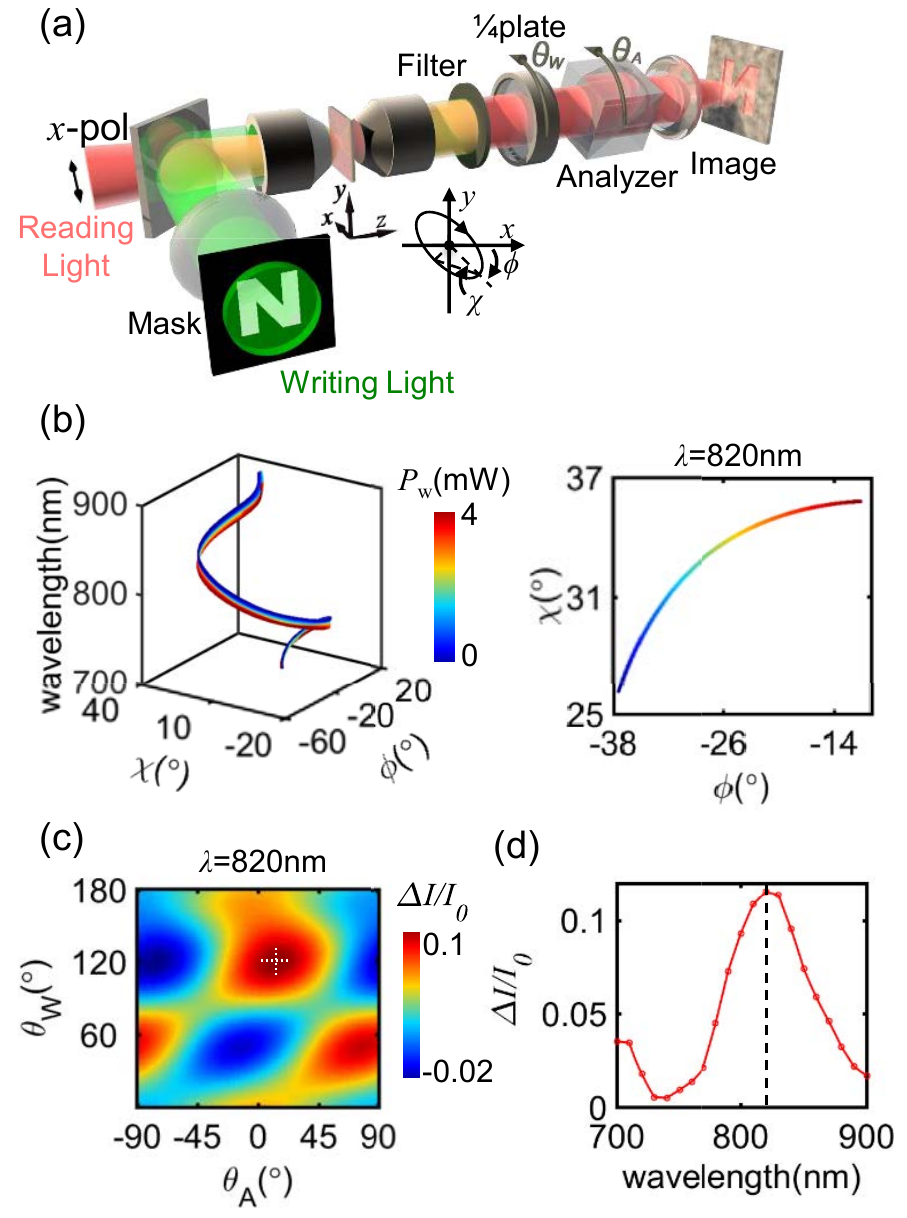} \caption{\label{fig2}
\textbf{Operation principle of MS-OASLM.} \textbf{a}, A schematic of the MS-OASLM imaging projection system. $x$-polarized reading light becomes elliptically polarized after passing through the metasurface, which are defined by azimuth rotation $\phi$ and ellipticity angle $\chi$. Positive values of $\phi$ and $\chi$ correspond to clockwise rotation of polarization azimuth and right-handed ellipse, as observed against propagation direction. A input mask is imaged onto the metasurface plane using 532~nm green light, and it is duplicated in form of a spatially inhomogeneous polarization distribution in readout light. Such polarization replica is transformed into a intensity replica by filtering through polarization elements. \textbf{b}, Nonlinear tuning of the reading light polarization as functions of wavelength and green light power. Under green light irradiation, the $\phi$-$\chi$ trajectory suffers obvious blue shift, i.e. the total curve shift to shorter wavelength end. The writing light power ($P_w$) is shown in the color bar. Right panel shows nonlinear change of polarization at 820~nm. \textbf{c}, Nonlinear intensity modulation transformed from polarization modulation using a combination of waveplate and polarizer, which is normalized to the input signal light power. $\theta_{W}$ and $\theta_{A}$ are azimuth angles of waveplate and analyzer, respectively. In the shown example, wavelength of the signal light is 820~nm. And the maximum modulation is indicated by a white cross. \textbf{d}, Maximum nonlinear intensity modulation for different reading light wavelengths, and 820~nm gives the strongest modulation magnitude.}
\end{figure*}

Recently, the emergence of metasurfaces provides new candidates for nonlinear modulating materials for OASLM. The metasurfaces could localize the light field within nanoscale dimensions, and significantly amplify optical energy density, thus boost the nonlinearity with orders of magnitude larger than natural materials.\cite{zheludev2012metamaterials,kauranen2012nonlinear,li2017nonlinear,krasavin2018free} Based on such advantages, significant nonlinear modulation over the light intensity, phase, and polarization under weak pump light have been demonstrated at nanoscales.\cite{macdonald2009ultrafast,nikolaenko2010carbon,wurtz2011designed,ren2011nanostructured,ren2012giant,ren2017reconfigurable} Furthermore, subwavelength sized building blocks of the metasurfaces not only guarantee the light transmission (or reflection) without diffraction, but also act as the natural pixels promising ultrahigh resolution inaccessible by the current SLM technologies. 

Here, we present a novel OASLM based on the metasurface framework (MS-OASLM). The operation principle of our device relies on giant nonlinear polarization control over the signal light by another pump light using the chiral metasurfaces. Both electrodes and photosensing layers are not needed at all, which provides the device with advantages of compact profile and easy fabrication. Our MS-OASLM works in an all-solid-state fashion by escaping from long lasting tradition of constructing the SLMs using the LCs.\cite{Efron1994Spatial,yang2014fundamentals} This not only avoids sophisticated rubbing and sealing for the LC assembly, but also makes the SLM applicable under rigid conditions, for example vacuum environment. Profiting from distinct ability of the metasurfaces in manipulating light behaviors at nanoscales, the MS-OASLM shows an advantage of ultra-thin thickness of only 400~nm, which is thinner than one-tenth of thickness of any conventional SLMs,\cite{Efron1994Spatial} and less than one-third of thickness of recently proposed LC based MS-EASLM.\cite{li2019phase} The subwavelength feature sizes in the metasurface render the MS-OASLM immune to the diffraction. Image projections by our MS-OASLM are demonstrated, which is fulfilled by transforming the polarization modulation into an intensity replica by polarizing elements. The resolution achieves about 250~lines per millimeter (lp/mm), which is more than one order of magnitude better than the commercial SLMs(Hamamatsu X10468-01, 25lp/mm). Such MS-OASLM would provide a flexible and compact platform for the next generation ultrahigh resolution optical displays, and all-optical applications including the parallel image processing and computation. 


The structure of the metasurface is shown in Fig.~1. It consists of a metallic nanostructure layer, which is covered by a 300~nm photoactive ethyl red azo polymer as the nonlinear switching layer. The building blocks of the nanostructure are composed of ``L''-shaped slits, each of which was fabricated via focused ion-beam (FIB) milling through a 100~nm thick gold film supported by a 500~$\mu$m thick fused quartz substrate. The nanostructure lattice constant is 300~nm, with an entire array footprint of 40$\times$40$\mu$m$^2$. Because the nanoblocks are chiral in geometry, an initially linearly polarized wave would become elliptically polarized with its azimuth rotated after passing through such medium. The polarization changes are characterized in terms of azimuth rotation $\phi$ and ellipticity angle $\chi$ (defined in Fig.~2(a)), and measured using a home-built polarimeter. Blue trajectory in Fig.~2(b) gives spectral dispersions of $\phi$ and $\chi$ in wavelength range of 700 to 900~nm. And as high as -37.43$^\circ$ in $\phi$ and 26.1$^\circ$ in $\chi$ are achieved for $x$-polarized incident light at 820~nm. Under external green light irradiation (532~nm), the azo molecules initially in trans state convert to cis state (as shown in Fig.~1). Such structural isomerization alters refractive index of the polymer, consequently changes resonance conditions of nanostructures. Consequently, a blue-shift of $\phi$-$\chi$ trajectory happens and the polarization states of the transmitted light are be actively varied under just milliwatt green light power ($P_w$), as indicated by Fig.~2(b). And as large as 26.3$^\circ$ and 9.7$^\circ$ nonlinear changes in $\phi$ and $\chi$ are observed at 820~nm, respectively.

The most widely application of SLM belongs to the image projections. To accomplish this goal, a MS-OASLM image projecting system was built, as illustrated in Fig.~2(a). A series of binary masks, which were made by milling transparent letters of ``I $\heartsuit$ N K U'' through an opaque metal film, were put in the green writing beam and imaged onto the metasurface plane by a combination of a lens ($f$=200~mm) and an objective (10$\times$, NA 0.25). In this way, the mask images were duplicated into the polymer layer in form of the spatial heterogeneity of the azo molecular structures, which would further transferred into the spatial polarization variance of the reading light. The output from a $x$-polarized tunable supercontinuum laser was used as the reading beam and impinged normally onto the metasurface. Another objective (10$\times$, NA 0.1) was put on the transmission side to collect the transmitted light, and a long pass filter to isolate the green light. A combination of a quarter waveplate and a polarizer was used to thus translate the polarization profile modulations into intensity distributions according to the Malus' law. For a given writing light power level, the nonlinear modulations of the intensity are dependent on the wavelength, azimuth angles of waveplate $\theta_{W}$ and analyzer $\theta_{A}$. In our experiment, the reading light was selected as 820~nm, meanwhile waveplate azimuth oriented along 120$^\circ$ and analyzer axis directed along 14$^\circ$, which is the condition for the largest normalized intensity modulation depth, as shown in Fig.~2(c) and (d). The signal beam was finally photographed onto a CMOS camera (Nikon, DS-2MBWc). 

The final images are given in Fig.~3(a), which present reasonable reproduction of the writing images. Furthermore, our MS-OASLM also functions as a wavelength converter here, which manages to transfer two dimensional information in visible (532~nm) to infrared (820~nm) channels in a full parallel fashion. Such wavelength converter is important to relocate optical channels between two separate optical systems. The spatial resolution is an important parameter of SLM. The higher resolution means the finer modulation over optical wavefront. The subwavelength feature sizes make the metasurface behave as a homogeneous film without optical diffraction. And ideally each unit cell acts as one pixel suggesting a high SLM resolution. To assess the spatial resolution, we replace the letter masks using resolution test charts, which consist of 2 sets of elements. Each element encloses three horizontal and three vertical lines. The charts are in photographic negative format fabricated by milling through an opaque metal film. The resolution test charts are written onto the metasurface plane using green light and form writing images with sizes ranging from 10 to 20~$\mu$m considering the magnification ratio of the optical system, as shown in first row of Fig.~3(b). The readout images by the infrared light are given in second row. The elements with sizes of 10~$\mu$m are well recognized as three distinct lines without any blurring into one another. This implies a spatial resolution of about 250~lp/mm for our OASLM, corresponding to about 2~$\mu$m for the single pixel size. Such resolution is much better than both previously reported LC based OASLMs \cite{white1970liquid,fukushima1991real,miri2014Liquid} and a typical commercial EASLM (Hamamatsu X10468, 25lp/mm). 

\begin{figure} [tp]
\includegraphics[width=90mm]{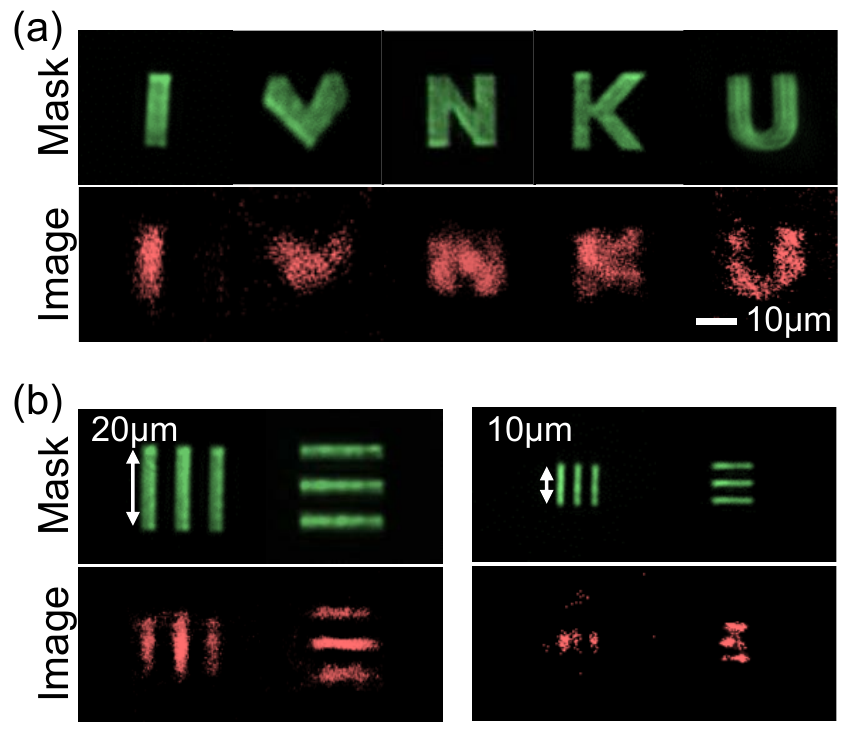} \caption{\label{fig3}
\textbf{Projection images of binary masks and resolution test charts} \textbf{a}, Binary masks include a series of transparent letters of ``I $\heartsuit$ N K U'' milled through an opaque metal film. First row gives optical images of the masks captured directly using the green light (532~nm). And images reproduced by infrared signal beam (820~nm) are shown in the second row. Scale bar is 10~$\mu$m. \textbf{b}, Images of different sized resolution test charts by green light and infrared beam are given in the first and second rows of each panel, respectively. The test charts include 2 sets of elements, each of which encloses three horizontal and three vertical lines. The sizes of chart images are labeled beside each patterns. The elements with sizes of 10~$\mu$m are well recognized as three distinct lines without any blurring into one another, corresponding to a spatial resolution of about 250~lp/mm.}
\end{figure} 

In conclusion, we demonstrate here a novel ultracompact OASLM based on the nonlinear metasurface. Benefiting from the outstanding light manipulation capacity at nanoscales of the metasurfaces, the MS-OASLM acquires a merit of unprecedented compactness with thickness of about 400~nm, which is less than one tenth of thickness of the traditional LC based EASLMs. And the subwavelength sized features in the metasurface render the MS-OASLM diffraction free leading to improved light utilization efficiency and suppressed diffraction noise in the readout light. In addition, because no electrode and photosensing layer is needed, the structure of the MS-OASLM is much simpler than the traditional OASLMs, thus significantly simplify the manufacturing process. We escape from the long lasting reliance on the LCs in manufacturing SLMs. This not only avoid sophisticated rubbing and sealing processes, but also endows the all-solid-state feature to the MS-OASLMs. Thus they become applicable in rigid environments, for example vacuum. And we demonstrate the application of image projection by our MS-OASLMs, whose resolution is shown to be one order of magnitude higher than the traditional SLMs devices. Our results would provide a new configuration for constructing the OASLMs. And the MS-OASLM would provide a great opportunities for novel optical displays and all-optical data processing techniques. 

\section{Acknowledgements}
This work was supported by National Key R\&D Program of China (2017YFA0305100, 2017YFA0303800); National Natural Science Foundation of China (61775106, 11711530205, 11374006, 11774185, 91750204, 11904182); the 111 Project (B07013); PCSIRT (IRT0149); the Open Research Program of Key Laboratory of 3D Micro/Nano Fabrication and Characterization of Zhejiang Province; the Fundamental Research Funds for the Central Universities; the Tianjin youth talent support program. We thank the Nanofabrication Platform of Nankai University for fabricating samples.



\section{References}
\begin{thebibliography}{10}
\newcommand{\enquote}[1]{``#1''}

\bibitem{goodman1968Introduction}
J.~W. Goodman, \emph{Introduction to Fourier optics} (Roberts \& Company,
  Colorado, 1968).

\bibitem{heynick1973special}
L.~Heynick, I.~Reingold, and A.~Sobel, \enquote{Special issue on display
  devices,} IEEE Trans. Electron Devices \textbf{20}, 917--920 (1973).

\bibitem{casasent1978optical}
D.~Casasent, \enquote{Optical data processing: applications,} in
  \enquote{Berlin and New York, Springer-Verlag (Topics in Applied Physics.
  Volume 23). 1978. 286 p (For individual items see A78-37895 to A78-37899),} ,
  vol.~23 (1978), vol.~23.

\bibitem{tanguay1983spatial}
A.~R. Tanguay~Jr, \enquote{Spatial light modulators: Critical issues,} Opt.
  Eng. \textbf{22} (1983).

\bibitem{Efron1994Spatial}
U.~Efron, \emph{Spatial light modulator technology: materials, devices, and
  applications} (Marcel Dekker, New York, 1994).

\bibitem{savage2009digital}
N.~Savage, \enquote{Digital spatial light modulators,} Nat. Photonics
  \textbf{3}, 170 (2009).

\bibitem{vettese2010microdisplays}
D.~Vettese, \enquote{Microdisplays: Liquid crystal on silicon,} Nat. Photonics
  \textbf{4}, 752 (2010).

\bibitem{maimone2017holographic}
A.~Maimone, A.~Georgiou, and J.~S. Kollin, \enquote{Holographic near-eye
  displays for virtual and augmented reality,} ACM T. Graphic \textbf{36}, 85
  (2017).

\bibitem{christian2013survey}
J.~A. Christian and S.~Cryan, \enquote{A survey of lidar technology and its use
  in spacecraft relative navigation,} in \enquote{AIAA Guidance, Navigation,
  and Control (GNC) Conference,}  (2013), p. 4641.

\bibitem{harke2008resolution}
B.~Harke, J.~Keller, C.~K. Ullal, V.~Westphal, A.~Sch{\"o}nle, and S.~W. Hell,
  \enquote{Resolution scaling in sted microscopy,} Opt. Express \textbf{16},
  4154--4162 (2008).

\bibitem{barredo2016atom}
D.~Barredo, S.~De~L{\'e}s{\'e}leuc, V.~Lienhard, T.~Lahaye, and A.~Browaeys,
  \enquote{An atom-by-atom assembler of defect-free arbitrary two-dimensional
  atomic arrays,} Science \textbf{354}, 1021--1023 (2016).

\bibitem{chiou2005massively}
P.~Y. Chiou, A.~T. Ohta, and M.~C. Wu, \enquote{Massively parallel manipulation
  of single cells and microparticles using optical images,} Nature
  \textbf{436}, 370 (2005).

\bibitem{beckers1993adaptive}
J.~M. Beckers, \enquote{Adaptive optics for astronomy: principles, performance,
  and applications,} Annu. Rev. Astron. Astrophys. \textbf{31}, 13--62 (1993).

\bibitem{katz2012looking}
O.~Katz, E.~Small, and Y.~Silberberg, \enquote{Looking around corners and
  through thin turbid layers in real time with scattered incoherent light,}
  Nat. Photonics \textbf{6}, 549 (2012).

\bibitem{moddel1997spatial}
G.~Moddel and P.~R. Barbier, \enquote{Spatial light modulators: Processing
  light in real time,} Opt. Photonics News \textbf{8}, 17--21 (1997).

\bibitem{yang2014fundamentals}
D.-K. Yang, \emph{Fundamentals of liquid crystal devices} (John Wiley \& Sons,
  2014).

\bibitem{margerum1970reversible}
J.~D. Margerum, J.~Nimoy, and S.~Y. Wong, \enquote{Reversible ultraviolet
  imaging with liquid crystals,} Appl. Phys. Lett. \textbf{17}, 51--53 (1970).

\bibitem{white1970liquid}
D.~L. White and M.~Feldman, \enquote{Liquid-crystal light valves,} Electron.
  Lett. \textbf{6}, 837--839 (1970).

\bibitem{shrestha2015high}
P.~K. Shrestha, Y.~T. Chun, and D.~Chu, \enquote{A high-resolution optically
  addressed spatial light modulator based on zno nanoparticles,} Light: Sci.
  Appl. \textbf{4}, e259 (2015).

\bibitem{miri2014Liquid}
K.~Miri~Gelbaor, K.~Matvey, L.~Victor, C.~Neil, and .~Abdulhalim, I.,
  \enquote{Liquid crystal high-resolution optically addressed spatial light
  modulator using a nanodimensional chalcogenide photosensor,} Opt. Lett.
  \textbf{39}, 2048--51 (2014).

\bibitem{smith1985lasers}
S.~Smith, \enquote{Lasers, nonlinear optics and optical computers,} Nature
  \textbf{316}, 319 (1985).

\bibitem{zhang1997incoherent}
J.~Zhang, H.~Wang, S.~Yoshikado, and T.~Aruga, \enquote{Incoherent-to-coherent
  conversion by use of the photorefractive fanning effect,} Opt. Lett.
  \textbf{22}, 1612--1614 (1997).

\bibitem{shih2001all}
M.~Shih, A.~Shishido, and I.~Khoo, \enquote{All-optical image processing by
  means of a photosensitive nonlinear liquid-crystal film: edge enhancement and
  image addition-subtraction,} Opt. Lett. \textbf{26}, 1140--1142 (2001).

\bibitem{woods2012optical}
D.~Woods and T.~J. Naughton, \enquote{Optical computing: Photonic neural
  networks,} Nat. Phys. \textbf{8}, 257 (2012).

\bibitem{solodar2016infrared}
A.~Solodar, T.~A. Kumar, G.~Sarusi, and I.~Abdulhalim, \enquote{Infrared to
  visible image up-conversion using optically addressed spatial light modulator
  utilizing liquid crystal and ingaas photodiodes,} Appl. Phys. Lett.
  \textbf{108}, 021103 (2016).

\bibitem{xu1995holographic}
J.~Xu, G.~Zhang, Q.~Wu, Y.~Liang, S.~Liu, Q.~Sun, X.~Chen, and Y.~Shen,
  \enquote{Holographic recording and light amplification in doped polymer
  film,} Opt. Lett. \textbf{20}, 504--506 (1995).

\bibitem{grunnet1997spontaneous}
A.~Grunnet-Jepsen, C.~L. Thompson, and W.~E. Moerner, \enquote{Spontaneous
  oscillation and self-pumped phase conjugation in a photorefractive polymer
  optical amplifier,} Science \textbf{277}, 549--552 (1997).

\bibitem{hesselink1998photorefractive}
L.~Hesselink, S.~S. Orlov, A.~Liu, A.~Akella, D.~Lande, and R.~R. Neurgaonkar,
  \enquote{Photorefractive materials for nonvolatile volume holographic data
  storage,} Science \textbf{282}, 1089--1094 (1998).

\bibitem{goonesekera2000image}
A.~Goonesekera, D.~Wright, and W.~E. Moerner, \enquote{Image amplification and
  novelty filtering with a photorefractive polymer,} Appl. Phys. Lett.
  \textbf{76}, 3358--3360 (2000).

\bibitem{yaqoob2008optical}
Z.~Yaqoob, D.~Psaltis, M.~S. Feld, and C.~Yang, \enquote{Optical phase
  conjugation for turbidity suppression in biological samples,} Nat. Photonics
  \textbf{2}, 110 (2008).

\bibitem{tay2008updatable}
S.~Tay, P.-A. Blanche, R.~Voorakaranam, A.~Tun{\c{c}}, W.~Lin, S.~Rokutanda,
  T.~Gu, D.~Flores, P.~Wang, G.~Li \emph{et~al.}, \enquote{An updatable
  holographic three-dimensional display,} Nature \textbf{451}, 694 (2008).

\bibitem{beard1973ac}
T.~Beard, W.~Bleha, and S.-Y. Wong, \enquote{ac liquid-crystal light valve,}
  Appl. Phys. Lett. \textbf{22}, 90--92 (1973).

\bibitem{moddel1989high}
G.~Moddel, K.~Johnson, W.~Li, R.~Rice, L.~Pagano-Stauffer, and M.~Handschy,
  \enquote{High-speed binary optically addressed spatial light modulator,}
  Appl. Phys. Lett. \textbf{55}, 537--539 (1989).

\bibitem{fukushima1990bistable}
S.~Fukushima, T.~Kurokawa, S.~Matsuo, and H.~Kozawaguchi, \enquote{Bistable
  spatial light modulator using a ferroelectric liquid crystal,} Opt. Lett.
  \textbf{15}, 285--287 (1990).

\bibitem{solodar2019highly}
A.~Solodar, H.~Manis-Levy, G.~Sarusi, and I.~Abdulhalim, \enquote{Highly
  sensitive liquid crystal optically addressed spatial light modulator for
  infrared-to-visible image up-conversion,} Opt. Lett. \textbf{44}, 1269--1272
  (2019).

\bibitem{li1995resolution}
W.~Li and G.~Moddel, \enquote{Resolution limits from charge transport in
  optically addressed spatial light modulators,} J. Appl. Phys. \textbf{78},
  6923--6935 (1995).

\bibitem{zheludev2012metamaterials}
N.~I. Zheludev and Y.~S. Kivshar, \enquote{From metamaterials to metadevices,}
  Nat. Mater. \textbf{11}, 917 (2012).

\bibitem{kauranen2012nonlinear}
M.~Kauranen and A.~V. Zayats, \enquote{Nonlinear plasmonics,} Nat. Photonics
  \textbf{6}, 737 (2012).

\bibitem{li2017nonlinear}
G.~Li, S.~Zhang, and T.~Zentgraf, \enquote{Nonlinear photonic metasurfaces,}
  Nat. Rev. Mater. \textbf{2}, 17010 (2017).

\bibitem{krasavin2018free}
A.~V. Krasavin, P.~Ginzburg, and A.~V. Zayats, \enquote{Free-electron optical
  nonlinearities in plasmonic nanostructures: A review of the hydrodynamic
  description,} Laser Photonics Rev. \textbf{12}, 1700082 (2018).

\bibitem{macdonald2009ultrafast}
K.~F. MacDonald, Z.~L. S{\'a}mson, M.~I. Stockman, and N.~I. Zheludev,
  \enquote{Ultrafast active plasmonics,} Nat. Photonics \textbf{3}, 55 (2009).

\bibitem{nikolaenko2010carbon}
A.~E. Nikolaenko, F.~De~Angelis, S.~A. Boden, N.~Papasimakis, P.~Ashburn,
  E.~Di~Fabrizio, and N.~I. Zheludev, \enquote{Carbon nanotubes in a photonic
  metamaterial,} Phys. Rev. Lett. \textbf{104}, 153902 (2010).

\bibitem{wurtz2011designed}
G.~A. Wurtz, R.~Pollard, W.~Hendren, G.~Wiederrecht, D.~Gosztola, V.~Podolskiy,
  and A.~V. Zayats, \enquote{Designed ultrafast optical nonlinearity in a
  plasmonic nanorod metamaterial enhanced by nonlocality,} Nat. Nanotechnol.
  \textbf{6}, 107 (2011).

\bibitem{ren2011nanostructured}
M.~Ren, B.~Jia, J.-Y. Ou, E.~Plum, J.~Zhang, K.~F. MacDonald, A.~E. Nikolaenko,
  J.~Xu, M.~Gu, and N.~I. Zheludev, \enquote{Nanostructured plasmonic medium
  for terahertz bandwidth all-optical switching,} Adv. Mater. \textbf{23},
  5540--5544 (2011).

\bibitem{ren2012giant}
M.~Ren, E.~Plum, J.~Xu, and N.~I. Zheludev, \enquote{Giant nonlinear optical
  activity in a plasmonic metamaterial,} Nat. Commun. \textbf{3}, 833 (2012).

\bibitem{ren2017reconfigurable}
M.-X. Ren, W.~Wu, W.~Cai, B.~Pi, X.-Z. Zhang, and J.-J. Xu,
  \enquote{Reconfigurable metasurfaces that enable light polarization control
  by light,} Light: Sci. Appl. \textbf{6}, e16254 (2017).

\bibitem{li2019phase}
S.-Q. Li, X.~Xu, R.~M. Veetil, V.~Valuckas, R.~Paniagua-Dom\'{\i}nguez, and
  A.~I. Kuznetsov, \enquote{Phase-only transmissive spatial light modulator
  based on tunable dielectric metasurface,} Science \textbf{364}, 1087--1090
  (2019).

\bibitem{fukushima1991real}
S.~Fukushima, T.~Kurokawa, and M.~Ohno, \enquote{Real-time hologram
  construction and reconstruction using a high-resolution spatial light
  modulator,} Appl. Phys. Lett. \textbf{58}, 787--789 (1991).

\end{thebibliography}
\end{document}